# Influence of Humidity, Temperature and Radicals on the Formation and Thermal Properties of Secondary Organic Aerosol (SOA) from Ozonolysis of β-pinene

*Eva U. Emanuelsson, Ågot K. Watne, Anna Lutz, Evert Ljungström and Mattias Hallquist\**

Atmospheric Science, Department of Chemistry and Molecular Biology, University of Gothenburg, SE-412 96 Gothenburg, Sweden

*Corresponding Author, e-mail: hallq@chem.gu.se, phone: +46 (0)31-786 9019

**Abstract**

The influence of water and radicals on SOAs produced by β-pinene ozonolysis was investigated at 298 and 288 K using a laminar flow reactor. A volatility tandem differential mobility analyzer (VTDMA) was used to measure the evaporation of the SOA, enabling the parameterization of its volatility properties. The parameters extracted included the temperature at which 50% of the aerosol had evaporated ($T_{VFR0.5}$) and the slope factor ($S_{VFR}$). An increase in $S_{VFR}$ indicates a broader distribution of vapor pressures for the aerosol constituents. Reducing the reaction temperature increased $S_{VFR}$ and decreased $T_{VFR0.5}$ under humid conditions but had less effect on $T_{VFR0.5}$ under dry conditions. In general, higher water concentrations gave lower $T_{VFR0.5}$ values, more negative $S_{VFR}$ values, and a reduction in total SOA production. The radical conditions were changed by introducing OH scavengers to generate systems with and without OH radicals and with different $[HO_2]/[RO_2]$ ratios. Presence of a scavenger and lower $[HO_2]/[RO_2]$ ratio reduced SOA production. Observed changes in $S_{VFR}$ values could be linked to the more complex chemistry that occurs in the absence of a scavenger, and indicated that additional $HO_2$ chemistry gives products with a wider range of vapor pressures. Updates to existing ozonolysis mechanisms with routes that describe the observed responses to water and radical conditions for monoterpenes with endocyclic and exocyclic double bonds are discussed.

**Introduction**

Atmospheric aerosol particles have adverse effects on both the global radiation budget and human health.[1] These effects depend on the particles' properties, including their number concentration, size distribution and chemical composition.[2] In most cases, the sub-micron fraction of the atmospheric aerosol is dominated by organic compounds.[3,4] Atmospheric oxidation of organic trace gases is a major source of low volatility organic compounds, which are converted to secondary organic aerosol (SOA) via gas-to-particle conversion.[5] The abundance of SOA is often higher than that of the primary organic aerosol (POA), and on the global scale, biogenic SOA is much more abundant than anthropogenic SOA.[6] Biogenic emissions of monoterpenes ($C_{10}H_{16}$) are an important source of SOA due to their considerable potential for SOA formation and their large-scale emission.[7,8]

SOA has a complex composition and is consequently difficult to accurately describe and quantify[9]. A measure of its volatility at a given temperature would be the sum of the actual equilibrium partial pressures of the components constituting the SOA particles.[10] In principle, this is a quantity that should be possible to calculate for use e.g. in atmospheric models.[11-13] However, the detailed composition of SOA particles is generally unknown, let alone the vapor pressures of the pure compounds.[14] Even if these quantities were known, it would be difficult to estimate the activity factors of each compound in the SOA mixture. A further complication is the non-equilibrium situation in the atmosphere, which drives the partitioning of compounds between the gas and condensed phase. Nevertheless, measurements of the integrated volatility of aerosol particles can be useful in understanding the processes that affect or are affected by changes in the chemical composition of the SOA as well as its molecular interactions and phase dynamics.[15] It is also valuable for the direct quantification of the volatility extremes, i.e. to distinguish between a non-volatile and a volatile aerosol particle.[16-18] Volatility measurements of SOA have largely been used to investigate the partitioning of products between the gaseous and condensed phases[19], and how SOA changes over time, i.e. as it ages.[15,20,21] A number of studies have used mass spectrometry to determine the volatility of a specific organic aerosol in order to investigate the relationship between chemical composition and volatility.[12,22-24] The low volatility fraction of SOA particles has been used to monitor the oligomerization processes that form atmospheric macromolecules.[25] Recently, there are also several studies suggesting that a low effective volatility depends more on the low mass transfer rates and less on the partitioning.[24,26,27]

The atmospheric oxidation of terpenes to form SOA is initiated by radicals (i.e. OH or $NO_3$) or ozone, with ozonolysis often being a dominant process.[28] The mechanisms by which ozone reacts with unsaturated hydrocarbons in the gas phase have been described in detail by Johnson and Marston.[29] Our current mechanistic knowledge regarding the ozonolysis of β-pinene has been summarized by various authors. Scheme 1 shows a simplified version of the mechanism presented by Nguyen et al.[30], with the addition of R3 a and R3 b as suggested by Winterhalter et al.[31]

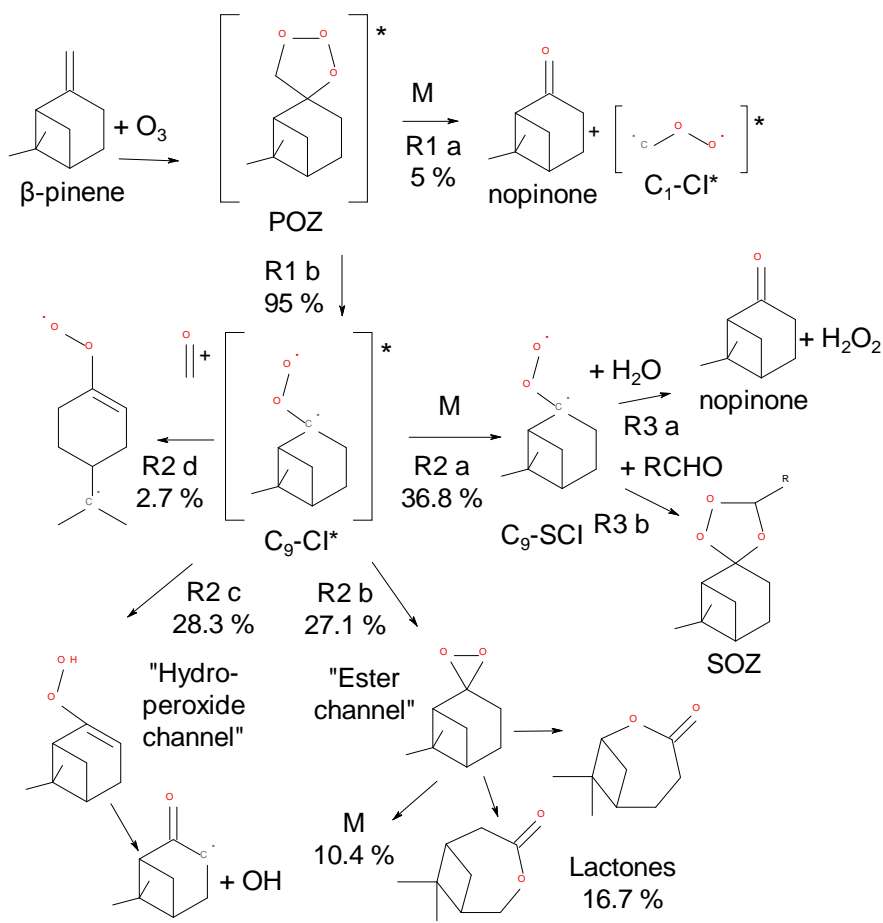

***Scheme 1.*** Major reaction paths in the ozonolysis of β-pinene. The routes and yields were taken from Nguyen et al.[30], omitting isomers and paths with yields below 1%. The reactions of the SCI (R3 a and R3 b) were suggested by Winterhalter et al.[31]

Reactions that are important for this work are discussed below; for more details, the reader is referred to the original study. Ozone adds to the double bond of the terpene to form a primary ozonide (POZ). This POZ is unstable and will decompose into a carbonyl and an excited biradical, the so-called Criegee intermediate (CI*). Two channels are then available depending on which of the two O-O bonds is broken. In the case of β-pinene, the POZ can form nopinone and the $C_1$-CI* radical ($CH_2OO$) (R1 a), but will predominantly generate HCHO and the $C_9$-CI* radical (R1 b). The CI* radical is both vibrationally and electronically excited when formed, since this reaction is highly exothermic.[32] The excited $C_9$-CI* will meet with one of several possible fates. It can isomerize via the ester channel (R2 b) to form a dioxirane that will subsequently decompose into various species including lactones. In the hydroperoxide channel (R2 c), the excited $C_9$-CI* decomposes into a hydroperoxide that in turn decomposes into hydroxyl and alkyl radicals. The alkyl radical reacts rapidly with molecular oxygen to form the first generation of peroxy radicals, which can subsequently participate in classical radical chain reactions.[33] Ring opening of the four-membered ring of the $C_9$-CI* radical (R2 d) is a minor pathway. The excited $C_9$-CI* can also be collisionally stabilized (R2 a) to generate a stabilized Criegee intermediate (SCI). $C_9$-SCI can react with water (R3 a) or carbonyl compounds (R3 b). The former will dominate due to the high concentration of water relative to organic carbonyls under atmospheric conditions.[31] Recent reinvestigations of the atmospheric importance of the $SO_2$ – SCI reaction, first suggested by Cox et al.[34], suggested that this process can compete with the water reaction for larger SCIs.[35,36]

One important aspect of the CI* hydroperoxide channel (R2 c) is that it produces OH radicals that will induce secondary chemical degradation. In the laboratory, an OH-scavenger such as 2-butanol or cyclohexane can be introduced to reduce the influence of OH chemistry.[37] However, this changes the radical distribution of the reaction mixture as well as removing OH, since the degradation of the OH-scavenger generates hydroperoxy ($HO_2$) and organic peroxy radicals ($RO_2$). Cyclohexane gives lower $HO_2/RO_2$ ratios than 2-butanol. This shift in radical distribution can be used to gain further knowledge of the ozonolysis reaction paths.[23,33,38,39]

Using the Gothenburg Flow Reactor for Oxidation Studies at low Temperatures (G-FROST) facility, Jonsson et al.[40] demonstrated that water influenced both the mass and number concentrations ($M_{10-300nm}$, $N_{10-300nm}$) of the SOAs derived from the ozonolysis of limonene, $\Delta^3$-carene and α-pinene. Subsequent studies investigated the influence of radical chemistry[41] and temperature dependence[42]. Based on the results obtained, it was suggested that monoterpenes with an endocyclic double bond behaved differently to exocyclic systems, in terms of both the effect of water and radical chemistry.[23,38] In order to experimentally establish these differences, the studies of Jonsson et al[40-42] have now been extended by investigating the effect of humidity, temperature, initial reaction rate, and radical conditions on the aerosol formed by ozonolysis of the monoterpene β-pinene having an exocyclic double bond. In addition, volatility measurements provided new information on chemically-induced changes in the aerosol's properties.

**Experimental Methods**

The G-FROST reactor can be used to perform detailed studies on the formation and properties of SOA. The set-up has previously been described in detail by Jonsson et al.[40-42] and Pathak et al.[43] and so only a short description is presented here. The vertical halocarbon wax-coated Pyrex® glass laminar flow reactor (191 cm long, i.d. 10 cm) is situated in a temperature-controlled housing. β-pinene was carried by the humidified bulk flow (with or without an OH scavenger) and entered at the top of the flow reactor. Ozone was added through an injector with a mixing plunger, ensuring rapid mixing. The total flow was 1.6 LPM, but only the central part (0.94 LPM) was conveyed through a sampling funnel for analysis. The average reaction time of the sampled flow was 240 s assuming a fully developed laminar flow profile. In the current study, G-FROST was modified to accommodate experiments under constant pressure, in order to avoid fluctuations in reaction times and concentrations due to variation in the atmospheric pressure. The pressure in the flow reactor was measured with a sensor (PTB110, Vaisala) and was maintained at 1030 ±0.01 hPa by dynamic throttling of the outflow. The relative humidity of the reacting flow was measured with a dew point meter (Optidew, Michelle Instruments LTD) in the outflow, and the temperature was continuously monitored at four heights on the surface of the flow reactor. The initial concentrations of β-pinene were measured using adsorption cartridge sampling with subsequent thermal desorption and GC-FID quantification. Ozone levels were measured before and after each experiment using a UV photometric $O_3$ analyzer (model 49C, Thermo Environmental Instruments Inc.). The precursor concentrations were not measured continuously to avoid damage to equipment and maintain constant flow rates.

The experimental conditions are summarized in Table 1. The experiments were labeled according to the β-pinene concentration used (L=low, I=intermediate or H=high), temperature (298 or 288K), and the use of a scavenger (no=no scavenger, but=2-butanol, cyc=cyclohexane).

Three β-pinene concentrations were considered: low ($2.0 \times 10^{12}$ molecules cm$^{-3}$), intermediate ($2.7 \times 10^{12}$ molecules cm$^{-3}$) and high ($4.1 \times 10^{12}$ molecules cm$^{-3}$). The concentrations of ozone in each case were selected to achieve the same initial reaction rate as used in previous studies.[40-42] In all experiments, the RH was changed in four steps, from 12% to 15%, 30%, and finally 50%. After each change in relative humidity, the system was allowed to equilibrate for at least two hours before further measurements were taken.

*Table 1.* Experimental conditions, including the initial concentration of the reactants. The experiments are named according to the concentration of β-pinene (L, I or H), temperature (288K or 298K) and OH-scavenger used (no, but or cyc). Within each experiment, the RH was changed in a stepwise fashion within the given interval. The statistical errors on the averages are given as one standard deviation for consecutive measurements.

| Experiment | Temperature (K) | OH scavenger | [a][β-pinene]$_0$ x 10$^{-12}$ (molecules cm$^{-3}$) | [b][O$_3$]$_0$ x 10$^{-13}$ (molecules cm$^{-3}$) | [c]reaction rates 10$^{-9}$ (molecules cm$^{-3}$ s$^{-1}$) | RH (%) |
|---|---|---|---|---|---|---|
| L288no | 288.4±0.1 | none | 1.96±0.05 | 4.16±0.01 | 1.57 | 11-47 |
| I288no | 288.5±0.1 | none | 2.71±0.11 | 4.16±0.01 | 2.17 | 12-49 |
| H288no | 288.5±0.1 | none | 4.09±0.08 | 4.16±0.01 | 3.27 | 13-53 |
| L288but | 288.5±0.1 | 2-butanol | 1.96±0.05 | 4.16±0.01 | 1.57 | 12-55 |
| I288but | 288.5±0.1 | 2-butanol | 2.71±0.11 | 4.16±0.01 | 2.17 | 12-56 |
| H288but | 288.5±0.1 | 2-butanol | 4.09±0.08 | 4.16±0.01 | 3.27 | 12-51 |
| L288cyc | 288.6±0.1 | cyclohexane | 1.96±0.05 | 4.16±0.01 | 1.57 | 12-51 |
| I288cyc | 288.6±0.1 | cyclohexane | 2.71±0.11 | 4.16±0.01 | 2.17 | 12-54 |
| H288cyc | 288.5±0.1 | cyclohexane | 4.09±0.08 | 4.16±0.01 | 3.27 | 11-51 |
| L298no | 298.4±0.1 | none | 1.97±0.05 | 3.78±0.01 | 1.84 | [d]10-38 |
| I298no | 298.2±0.1 | none | 2.72±0.11 | 3.78±0.01 | 2.54 | [d]10-39 |
| H298no | 298.2±0.1 | none | 4.10±0.08 | 3.78±0.01 | 3.28 | 12-51 |
| L298but | 298.2±0.1 | 2-butanol | 1.97±0.05 | 3.78±0.01 | 1.84 | [d]12-48 |
| I298but | 298.2±0.1 | 2-butanol | 2.72±0.11 | 3.78±0.01 | 2.54 | [d]10-54 |
| H298but | 298.2±0.1 | 2-butanol | 4.10±0.08 | 3.78±0.01 | 3.28 | [d]12-49 |
| L298cyc | 298.2±0.1 | cyclohexane | 1.97±0.05 | 3.78±0.01 | 1.84 | 13- 43 |
| I298cyc | 298.2±0.1 | cyclohexane | 2.72±0.11 | 3.78±0.01 | 2.54 | 12- 49 |
| H298cyc | 298.2±0.1 | cyclohexane | 4.10±0.08 | 3.78±0.01 | 3.28 | 12- 50 |

[a] Corresponding concentrations of β-pinene were ~164 ppb, ~ 109 ppb and ~79 ppb at 298 K.
[b] Corresponding concentration of ozone was ~1600 ppb at 298 K.
[c] The rate coefficient expression used is $1.74 * 10^{-15} * e^{-1297/T}$ cm$^3$ molecule$^{-1}$ s$^{-1}$.[44]
[d] RH is estimated based on earlier experiments due to measurement problems.

SOA mass and number concentrations were measured using a scanning mobility particle sizer (SMPS3936L75, TSI). The total particle number ($N_{10-300nm}$, # cm$^{-3}$) and mass ($M_{10-300nm}$, µg m$^{-3}$) concentrations were obtained from the average of ~ 2 hours of consecutive SMPS scans, with each scan lasting for 5 min. The SOA mass was calculated using a SOA density of 1.4 g cm$^{-3}$ as suggested by Hallquist et al.[6] The thermal properties of the SOA were characterized at 383K and for selected evaporation temperatures (298-483 K) using a volatility tandem differential mobility analyzer (VTDMA) setup. The VTDMA has been described in detail elsewhere.[10,20] Briefly, it consists of a size-selecting differential mobility analyzer (DMA) that is typically set at 60-75 nm, an oven unit where the evaporation occurs, charcoal denuders, and a second DMA together with a condensation particle counter (CPC3775, TSI) to characterize the resulting aerosol. Before the particles entered the VTDMA, they were dried using a Nafion® dryer (Perma Pure PD100T-12MSS) to reduce the potential effect of adsorbed water. The VTDMA sample flow was 0.3 SLPM, giving a median residence time of 2.0 s in the heated part of the oven, assuming a fully developed laminar flow. The volume fraction remaining, VFR(T), was derived (assuming spherical particles) from the measured particle modal diameter ($D_p$) at a temperature T, normalized to the selected diameter ($D_{pRef}$). The VFR at 383 K (VFR(383K)) was continuously monitored during all experiments except when detailed thermal characterisation was done at the highest and lowest humidity conditions. The VFR was then measured between 298 K and 483 K at intervals of 10-15 K. To obtain more information over the full range of evaporation temperatures, a sigmoidal curve was fitted to the full range of VFR data versus evaporation temperature.[45]

$$VFR_T = VFR_{max} + \frac{(VFR_{min} - VFR_{max})}{1 + \left(\frac{T_{position}}{T}\right)^{S_{VFR}}} \quad (1)$$

The $VFR_{max}$ and $VFR_{min}$ set the boundaries of the highest and lowest VFRs, respectively. The slope factor $S_{VFR}$ and the $T_{position}$ define the shape of the curve. The mid-position, $T_{position}$, sets the temperature at which $VFR_T$ is at $(VFR_{max} + VFR_{min})/2$. In order to more strictly define the most volatile and the non-volatile fraction, equation 1 was used to derive $VFR_{298K}$ and $VFR_{523K}$. In addition, the temperature at which the VFR is 0.5 ($T_{VFR0.5}$) was calculated. The $T_{VFR0.5}$ is a general measure of the SOA volatility: an increase in $T_{VFR0.5}$ corresponds to a less volatile aerosol particle. The slope factor $S_{VFR}$ measures the distribution of the volatility of the individual components of the particles: the major constituents of an aerosol with a relatively steep slope have a relatively narrow vapor pressure distribution.

**Results**

Table 1 shows the experimental conditions used for each experiment, and the results obtained are summarised in Table 2 with corresponding statistical errors at one standard deviation. The statistical errors in VFR(383K) are below 2%, which is in line with the uncertainties stated in our previous volatility measurement.[20] The effect of changes in humidity, concentration, and scavenger on aerosol mass, number, $S_{VFR}$, $T_{VFR0.5}$ and $VFR_{298K}$ are summarized in Table 3.

*Humidity effects*

Figure 1 shows how the mass and number concentrations as well as VFR(383K) varied with induced changes in relative humidity during a typical experiment. Figure 2 shows how the SOA number ($N_{10-300nm}$) (2a) and mass ($M_{10-300nm}$) (2b) concentrations decreased with increasing RH for six of the experiments.

*Table 2.* Summary of the experimental results. Mass and number concentrations of the formed SOA, with corresponding VFR(383K) values for each RH step examined in the experiments. At the highest and lowest relative humidities, the volatility was characterized and the parameters derived from the sigmoidal fit to these data are given (i.e. $S_{VFR}$, $T_{VFR0.5}$ and $VFR_{298K}$). The statistical errors on the averages are given as one standard deviation for consecutive measurements. $VFR_{298}$ is less than unity for experiments at 298K solely because of dilution.

| | Experimental results | | | | Results from sigmoidal fit | | |
|---|---|---|---|---|---|---|---|
| Experiment Index | RH (%) | Number (#/cm3) *10^4 | [a]Mass (µg m^-3) | VFR(383K) | $S_{VFR}$ | $T_{VFR0.5}$ (K) | $VFR_{298K}$ |
| L288no | 47 ± 0.1 | 7.7 ± 0.2 | 11.0 ± 0.4 | 0.181±0.002 | | | |
| | 24.9 ± 0.1 | 15.3 ± 0.3 | 16.5 ± 0.4 | 0.212±0.003 | | | |
| | 13.9 ± 0.1 | 23.2 ± 0.5 | 19.9 ± 0.8 | 0.241±0.003 | | | |
| | 11 ± 0.7 | 23.3 ± 1.2 | 20.3 ± 0.7 | 0.245±0.004 | -16.37±0.79 | 342.30±1.13 | 0.704±0.143 |
| | 13.6 ± 0.1 | 18.6 ± 0.3 | 18.1 ± 0.5 | 0.230±0.003 | | | |
| | 46.3 ± 0.1 | 6.8 ± 0.2 | 11.3 ± 0.3 | 0.177±0.002 | -15.68±0.46 | 351.12±0.71 | 0.744±0.110 |
| I288no | 49.4 ± 0.7 | 19.9 ± 1.7 | 29.3 ± 1.5 | 0.170±0.006 | -16.22±0.66 | 336.94±1.09 | 0.643±0.133 |
| | 29.2 ± 1.7 | 33.4 ± 0.9 | 35.7 ± 0.6 | 0.197±0.003 | | | |
| | 16.3 ± 0.3 | 47.9 ± 0.8 | 43.7 ± 1.1 | 0.251±0.105 | | | |
| | 12.1 ± 0.2 | 52.5 ± 0.5 | 45.8 ± 1.6 | 0.269±0.002 | -15.24±0.40 | 343.31±0.75 | 0.652±0.106 |
| | 16 ± 0.2 | 44.5 ± 0.4 | 41.5 ± 1.7 | 0.262±0.003 | | | |
| H288no | 53.1 ± 0.8 | 36.8 ± 2.9 | 63.4 ± 9.3 | 0.206±0.002 | -15.58±0.50 | 342.21±0.83 | 0.667±0.117 |
| | 17.1 ± 0.4 | 75.5 ± 0.7 | 88.1 ± 1.7 | 0.251±0.003 | | | |
| | 12.8 ± 0.2 | 77.2 ± 1.2 | 92.0 ± 2.2 | 0.258±0.003 | -14.41±0.45 | 349.28±0.82 | 0.712±0.117 |
| | 16.4 ± 0.2 | 68.1 ± 2.1 | 87.0 ± 2.9 | 0.250±0.008 | | | |
| L288but | 54 ± 0.1 | 0.34 ± 0.02 | 0.10 ± 0.01 | b.d. | | | |
| | 16.1 ± 0.1 | 1.26 ± 0.05 | 0.65 ± 0.04 | b.d. | | | |
| | 12.2 ± 0.1 | 1.66 ± 0.06 | 0.99 ± 0.07 | b.d. | | | |
| | 15.6 ± 0.1 | 1.18 ± 0.04 | 0.85 ± 0.07 | b.d. | | | |
| | 28.7 ± 0.1 | 0.58 ± 0.02 | 0.41 ± 0.06 | b.d. | | | |
| | 55 ± 0.2 | 0.27 ± 0.02 | 0.13 ± 0.01 | b.d. | | | |
| I288but | 55.8 ± 0.4 | 3.9 ± 0.1 | 3.7 ± 0.1 | 0.250±0.004 | -16.86±0.81 | 346.98±1.12 | 0.688±0.138 |
| | 27.8 ± 0.8 | 16.3 ± 0.7 | 13.3 ± 0.4 | 0.279±0.007 | | | |
| | 15.6 ± 0.4 | 25.7 ± 0.7 | 17.7 ± 0.7 | 0.313±0.005 | | | |
| | 12 ± 0.4 | 21.3 ± 0.6 | 14.9 ± 0.3 | 0.261±0.004 | -15.60±0.72 | 347.51±1.26 | 0.675±0.138 |
| | 15.4 ± 0.3 | 17.7 ± 0.4 | 13.2 ± 0.2 | 0.243±0.004 | | | |
| | 48.3 ± 0.9 | 4.7 ± 0.4 | 5.0 ± 0.3 | 0.205±0.002 | -17.02±0.77 | 341.38±1.28 | 0.648±0.134 |
| H288but | 51.3 ± 0.7 | 17.7 ± 0.5 | 22.8 ± 0.4 | 0.204±0.002 | -16.27±1.13 | 341.47±1.81 | 0.661±0.169 |
| | 28.4 ± 0.9 | 36.3 ± 0.7 | 32.8 ± 0.4 | 0.225±0.005 | | | |
| | 16.1 ± 0.4 | 50.6 ± 0.6 | 40.2 ± 0.6 | 0.247±0.002 | | | |
| | 11.9 ± 0.3 | 53.7 ± 0.7 | 42.9 ± 0.8 | 0.259±0.003 | -14.66±0.68 | 346.59±1.25 | 0.683±0.141 |
| | 15.1 ± 0.3 | 47.0 ± 0.4 | 39.2 ± 0.6 | 0.248±0.009 | | | |
| | 45.3 ± 1.1 | 19.1 ± 0.7 | 25.2 ± 1.2 | 0.204±0.003 | -16.21±0.66 | 341.33±0.98 | 0.673±0.129 |
| L288cyc | 51.4 ± 0.4 | 1.5 ± 0.1 | 1.4 ± 0.1 | 0.227±0.005 | -18.76±0.89 | 349.58±0.98 | 0.676±0.132 |
| | 16.1 ± 0.3 | 9.4 ± 0.3 | 7.0 ± 0.1 | 0.279±0.005 | | | |
| | 11.6 ± 0.4 | 9.8 ± 0.3 | 8.2 ± 0.3 | 0.275±0.006 | -16.34±0.50 | 356.90±0.66 | 0.748±0.109 |
| | 15.5 ± 0.3 | 7.3 ± 0.3 | 5.7 ± 0.3 | 0.267±0.005 | | | |
| | 27.8 ± 0.5 | 2.9 ± 0.1 | 3.1 ± 0.2 | 0.233±0.005 | | | |
| I288cyc | 53.6 ± 0.4 | 5.0 ± 0.5 | 9.6 ± 0.5 | 0.189±0.003 | -18.20±0.63 | 346.68±0.74 | 0.703±0.115 |
| | 16 ± 0.3 | 24.3 ± 0.6 | 22.2 ± 0.4 | 0.279±0.008 | | | |
| | 11.6 ± 0.3 | 28.4 ± 0.5 | 25.4 ± 0.5 | 0.265±0.004 | -16.54±0.32 | 354.89±0.41 | 0.745±0.083 |
| | 15.2 ± 0.3 | 24.8 ± 0.7 | 22.5 ± 0.4 | 0.260±0.005 | | | |
| | 27.8 ± 0.5 | 13.9 ± 1.3 | 15.8 ± 1.0 | 0.235±0.004 | | | |
| H288cyc | 47 ± 0.4 | 21.0 ± 0.6 | 35.8 ± 1.2 | 0.197±0.004 | -17.10±0.58 | 342.13±0.80 | 0.677±0.118 |
| | 27.4 ± 0.3 | 37.4 ± 0.9 | 46.9 ± 0.7 | 0.241±0.005 | | | |
| | 15.4 ± 0.4 | 54.1 ± 1.2 | 56.2 ± 2.0 | 0.261±0.005 | | | |
| | 11.4 ± 0.3 | 60.0 ± 0.9 | 56.1 ± 0.7 | 0.251±0.009 | -16.07±0.29 | 354.67±0.43 | 0.753±0.086 |
| | 51.2 ± 6.5 | 20.8 ± 5.6 | 35.9 ± 5.0 | 0.249±0.005 | -18.97±0.97 | 346.49±1.06 | 0.685±0.139 |
| L298no | [b]37.6 | 0.95 ± 0.05 | 0.7 ± 0.1 | 0.211±0.002 | | | |
| | [b]24.2 | 1.60 ± 0.06 | 1.2 ± 0.1 | 0.207±0.006 | | | |
| | [b]12.5 | 3.02 ± 0.07 | 2.6 ± 0.1 | 0.208±0.003 | | | |
| | [b]10.4 | 3.68 ± 0.09 | 3.5 ± 0.1 | 0.210±0.003 | -16.40±0.65 | 338.17±0.94 | 0.702±0.133 |
| | [b]12.8 | 3.17 ± 0.11 | 3.4 ± 0.2 | 0.213±0.003 | | | |
| | [b]38.7 | 1.45 ± 0.07 | 1.9 ± 0.2 | 0.205±0.003 | -16.76±1.01 | 345.63±1.35 | 0.694±0.156 |
| I298no | [b]39.3 | 3.8 ± 0.1 | 8.3 ± 0.4 | 0.200±0.001 | | | |
| | [b]23.0 | 6.0 ± 0.2 | 12.2 ± 0.3 | 0.216±0.001 | | | |
| | [b]12.6 | 10.9 ± 0.3 | 19.5 ± 0.4 | 0.225±0.001 | | | |
| | [b]10.3 | 12.3 ± 0.4 | 22.2 ± 0.5 | 0.226±0.001 | -17.33±0.35 | 350.33±0.46 | 0.739±0.088 |
| | [b]12.3 | 9.9 ± 0.1 | 19.2 ± 0.2 | 0.227±0.001 | | | |
| | [b]38.2 | 3.5 ± 0.1 | 8.7 ± 0.2 | 0.206±0.001 | -17.28±0.69 | 347.35±0.84 | 0.720±0.125 |
| H298no | 50.5 ± 0.1 | 8.1 ± 0.5 | 29.8 ± 0.7 | 0.195±0.003 | | | |
| | 15.4 ± 0.1 | 24.5 ± 0.3 | 51.8 ± 0.6 | 0.224±0.001 | | | |
| | 11.9 ± 0.3 | 26.8 ± 0.6 | 56.3 ± 0.4 | 0.227±0.000 | -16.82±0.29 | 350.16±0.40 | 0.736±0.083 |
| | [b]27.0 | 13.6 ± 0.2 | 39.1 ± 0.7 | 0.212±0.001 | | | |
| | [b]51.2 | 8.4 ± 0.1 | 31.3 ± 0.3 | 0.196±0.001 | -16.59±0.57 | 345.79±0.74 | 0.726±0.118 |
| L298but | [b]46.9 | 0.027 ± 0.009 | 0.0027 ± 0.0021 | b.d. | | | |
| | [b]29.2 | 0.044 ± 0.011 | 0.0052 ± 0.0010 | b.d. | | | |
| | [b]15.8 | 0.108 ± 0.012 | 0.0123 ± 0.0030 | b.d. | | | |
| | [b]12.3 | 0.132 ± 0.016 | 0.0133 ± 0.0025 | b.d. | | | |
| | [b]15.7 | 0.104 ± 0.016 | 0.0112 ± 0.0019 | b.d. | | | |
| | [b]48.0 | 0.058 ± 0.010 | 0.0032 ± 0.0004 | b.d. | | | |
| I298but | [b]54.1 | 0.57 ± 0.04 | 0.30 ± 0.02 | 0.239±0.011 | | | |
| | [b]32.2 | 0.81 ± 0.04 | 0.52 ± 0.03 | 0.260±0.011 | | | |
| | [b]15.1 | 1.45 ± 0.06 | 1.08 ± 0.06 | 0.246±0.012 | | | |
| | [b]10.5 | 1.82 ± 0.14 | 1.50 ± 0.12 | 0.237±0.008 | | | |
| | [b]13.7 | 1.55 ± 0.04 | 1.45 ± 0.06 | 0.250±0.008 | | | |
| | [b]52.6 | 0.60 ± 0.04 | 0.37 ± 0.03 | 0.235±0.011 | | | |
| H298but | [b]47.9 | 3.6 ± 0.1 | 9.0 ± 0.3 | 0.210±0.002 | | | |
| | [b]29.9 | 7.0 ± 0.2 | 15.5 ± 0.7 | 0.233±0.002 | | | |
| | [b]15.9 | 11.9 ± 0.2 | 21.6 ± 0.3 | 0.244±0.001 | | | |
| | [b]12.3 | 13.6 ± 0.4 | 25.2 ± 0.5 | 0.246±0.001 | -17.00±0.36 | 353.07±0.44 | 0.748±0.090 |
| | [b]48.5 | 3.6 ± 0.1 | 9.5 ± 0.3 | 0.208±0.002 | -17.05±0.75 | 347.19±0.94 | 0.717±0.131 |
| L298cyc | 31 ± 0.2 | 0.39 ± 0.03 | 0.15 ± 0.01 | 0.306±0.032 | | | |
| | 16.1 ± 0.1 | 0.91 ± 0.04 | 0.49 ± 0.02 | 0.299±0.020 | | | |
| | 12.6 ± 0.3 | 1.35 ± 0.06 | 0.88 ± 0.04 | 0.282±0.015 | | | |
| | 15.6 ± 0.2 | 0.98 ± 0.03 | 0.58 ± 0.03 | 0.295±0.015 | | | |
| | 43.4 ± 0.3 | 0.281±0.028 | 0.15 ± 0.01 | | | | |
| I298cyc | 30.9 ± 0.2 | 2.4 ± 0.1 | 4.3 ± 0.1 | 0.264±0.002 | | | |
|  | 15.5 ± 0.2 | 6.0 ± 0.1 | 10.7 ± 0.2 | 0.269±0.001 |  |  |  |
|  | 12.3 ± 0.3 | 7.1 ± 0.1 | 12.7 ± 0.3 | 0.266±0.001 | -17.59±0.51 | 356.80±0.65 | 0.792±0.106 |
|  | 15.5 ± 0.3 | 5.1 ± 0.1 | 8.6 ± 0.2 | 0.262±0.001 |  |  |  |
|  | 49.1 ± 0.3 | 1.1 ± 0.1 | 1.7 ± 0.1 | 0.235±0.003 | -18.27±1.27 | 353.81±1.36 | 0.744±0.164 |
| H298cyc | 49.7 ± 0.2 | 4.1 ± 0.1 | 15.7 ± 0.4 | 0.231±0.026 |  |  |  |
|  | 30.5 ± 0.2 | 8.0 ± 0.1 | 23.3 ± 0.4 | 0.251±0.012 |  |  |  |
|  | 15.7 ± 0.4 | 16.0 ± 0.2 | 33.4 ± 0.6 | 0.258±0.012 |  |  |  |
|  | 12.1 ± 0.2 | 20.0 ± 0.3 | 38.1 ± 0.4 | 0.243±0.016 | -17.11±0.86 | 354.65±1.10 | 0.777±0.140 |
|  | 15 ± 0.1 | 16.5 ± 0.2 | 34.6 ± 0.5 | 0.231±0.203 |  |  |  |
|  | [b]43.5 | 4.2 ± 0.1 | 17.6 ± 0.5 | 0.199±0.019 | -18.26±0.85 | 350.40±0.93 | 0.718±0.132 |

[a] Assumed SOA density of 1.4 g cm$^{-3}$
[b] RH is estimated based on earlier experiments due to measurement problems.
b.d. = below detection level

**Table 3.** Effect of changes in humidity, concentration, temperature and scavenger conditions on SOA mass, number, $S_{VFR}$, $T_{VFR0.5}$ and $VFR_{298K}$. An increase in more than 75% of the experiments is indicated with ++, an increase in more than 50% is indicated with +, a decrease in more than 75% of the experiments is indicated with --, a decrease in more than 50% of the experiments is indicated with -, and less than 50% increase or decrease is noted with 0. Within brackets the number of experiment with the dominant change and the total number of experiments are given.

| Compared conditions | Mass | Number | $S_{VFR}$ | $T_{VFR0.5}$ | $VFR_{298K}$ |
|---|---|---|---|---|---|
| **RH** (from 12 to 50%) | -- (14/14) | -- (14/14) | -- (11/14) | -- (12/14) | -- (13/14) |
| **Concentration** (from intermediate to high) | ++ (10/10) | ++ (10/10) | ++ (10/10) | + (6/10) | 0 (5/10) |
| **Temperature** (from 288 to 298K) | -- (12/12) | -- (12/12) | -- (11/12) | - (8/12) | ++ (10/12) |
| **Scavenger** (addition of scavenger) | -- (16/16) | -- (16/16) | -- (15/16) | ++ (13/16) | ++ (11/16) |
| **Scavenger** (change from but to cyc) | ++ (6/6) | ++ (6/6) | --(6/6) | ++ (5/6) | ++ (6/6) |

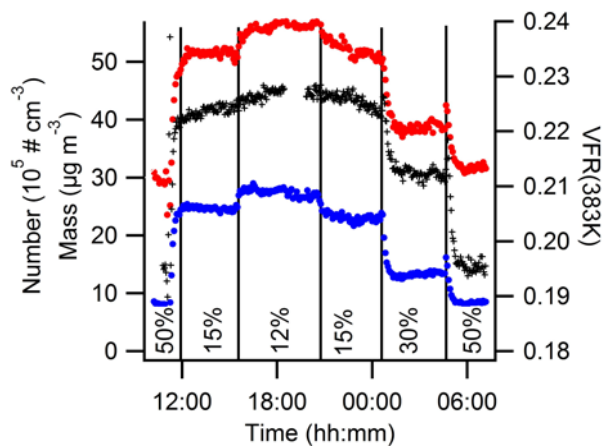

**Figure 1.** VFR(383K) (black), SOA number (blue) and mass (red) concentrations at different relative humidities in the H298no experiment. The number concentrations have been multiplied by $1 \times 10^{-5}$.

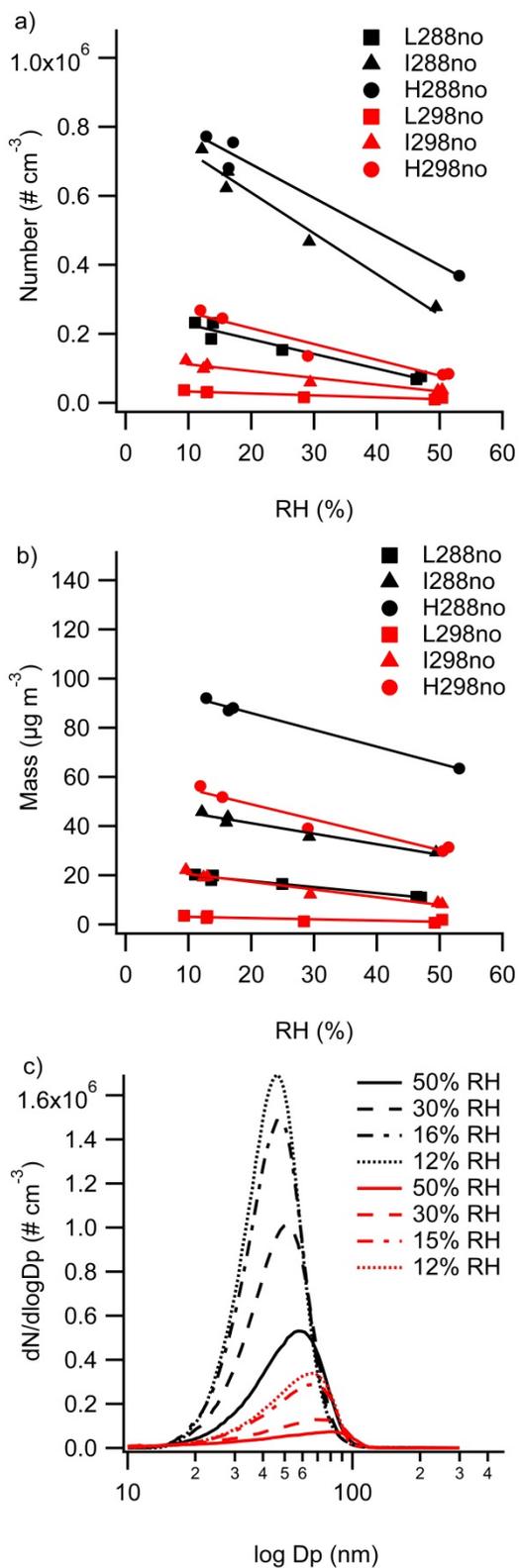

**Figure 2.** Effect of relative humidity on a) particle number, b) mass concentration and c) particle number size distributions at reaction temperatures of 288 K (black) and 298 K (red). The legends refer to the experimental conditions described in Table 1.

For all experiments, $M_{10-300nm}$ and $N_{10-300nm}$ decreased as the RH increased, i.e. there was a negative humidity effect. The magnitude of this effect varies between the systems. The $N_{10-300nm}$ was between 2.1 and 6.7 times higher in the dry experiments compared to the corresponding humid experiments. At 288 K, the humidity effect on $N_{10-300nm}$ decreased with the initial rate of reaction (increased concentration of β-pinene), and the effect of humidity was more pronounced when using an OH scavenger. This was also observed at 298 K, with exception of the experiments with the lowest produced aerosol mass. The influence of temperature on the humidity effect for $N_{10-300nm}$ depends on the initial concentration of β-pinene, i.e. the initial rate of reaction. For high concentration experiments, the relative change in $N_{10-300}$ due to humidity was always stronger at 298 K than at 288 K. The opposite was true in the low concentration experiments, but at the intermediate concentration there was no difference between the two temperatures. Because of the humidity effect, the $M_{10-300nm}$ value under dry conditions was between 1.4 and 9.5 times higher than that under humid conditions. This effect responded to changes in other parameters in the same way as for $N_{10-300nm}$, i.e. the effect decreased as the β-pinene concentration increased and was less pronounced in the absence of a scavenger.

For all conditions considered, the negative humidity effect on β-pinene SOA formation was different to that observed in previous studies on endocyclic terpene ozonolysis using G-FROST. Increases in humidity always increased the SOA mass in experiments using the monoterpenes α-pinene, $\Delta^3$-carene and limonene. However, the direction of the humidity effect on particle number depended on other parameters such as the temperature and the scavenger used. Our observations for β-pinene are consistent with at least two previous reports of a negative humidity effect on SOA formation.[23,46] However, a recent publication by von Hessberg et al.[47] presented aerosol yields for experiments without an OH scavenger at 293 K in which the yields under dry conditions were slightly lower than those obtained under humid conditions. The reason for this is not clear to us, but it should be noted that the water effect in that study was temperature-dependent and that at 273 K, the reported aerosol yield was highest under dry conditions, which is in keeping with our findings.[47]

*Temperature effects*

For all conditions, the experiments conducted at low temperature (288 K) produced a greater mass and number of particles than the corresponding experiments at 298 K, i.e. there was a negative temperature effect. Figures 2a and b illustrate the effect of temperature on $N_{10-300nm}$ and $M_{10-300nm}$ in experiments without an OH-scavenger. Figure 2c shows the size distributions for 298 and 288 K. As expected, the lower temperature promotes nucleation, producing a much higher peak than that achieved at the higher temperature. Despite the greater mass of particles at 288 K, the peak at this temperature is shifted slightly towards smaller diameter particles. The increase in SOA mass at lower temperatures is in agreement with the results of von Hessberg et al.[47] for dry conditions, but they reported an opposing aerosol mass - temperature relationship for humid conditions. Pathak et al.[48] reported an increase in aerosol mass with decreasing temperature that is consistent with our findings. Specifically, Pathak et al. found that reducing the temperature by 10 K increased the particle mass concentration by up to a factor of 2. The increase observed in our study was comparable or somewhat greater: the $M_{10-300nm}$ value for the high concentration 288 K experiments was between 1.5 and 2.7 times greater than that observed at 298 K. For the intermediate and low concentration experiments, the temperature effect was significantly stronger. The negative temperature effect on $M_{10-300nm}$ was always slightly stronger in the high humidity experiments than in those at low humidity.

Pathak et al.[48] attributed the temperature effect to the partitioning of products rather than changes in chemistry. This is in line with our observation that experiments with higher $M_{10-}$ $_{300nm}$ values had a less pronounced temperature effect, i.e. the relationship between SOA yield and organic aerosol mass is proportional to organic mass at low concentrations, and independent of organic mass at high concentrations.[6,49,50] A further complication is that reaction rates are set by the absolute water concentration and not the relative humidity. Thus, a secondary effect of temperature is that as the temperature falls, a given relative humidity value corresponds to a lower absolute water concentration. This will then make the rate of any reactions with water slower if their rate coefficients do not have a strong negative temperature dependence.

*Radical effects*

The measured SOA mass and number concentrations for the ozonolysis of β-pinene always increased in the following order: no scavenger >> cyclohexane > 2-butanol. This was true for all conditions, i.e. regardless of the RH, temperature, and initial rate. In our previous study on the ozonolysis of α-pinene, $\Delta^3$-carene and limonene, the use of 2-butanol as OH scavenger gave higher $M_{10-300nm}$ values whereas the opposite was observed in this work.[41] In other previous studies, the different scavenger effects seen with α- and β-pinene were attributed to the position of the double bond, with an exocyclic double bond producing less SOA as the $HO_2/RO_2$ ratio increased.[38,39] Our results from this study, which was performed with the same experimental set-up as that used by Jonsson et al.[41,42], confirm this observation. The scavenger effect on $M_{10-300nm}$ was always stronger at lower β-pinene concentrations. In addition, increasing the RH enhanced the scavenger effect at 288 K and also at 298 K in the high concentration experiments.

*Volatility*

Volatility changes were monitored at a fixed evaporation temperature of 383 K. Figure 1 shows how the VFR(383K) shifts due to changes in relative humidity. In all cases,

VFR(383K) decreased, indicating an increase in particle volatility, as the water concentration increased. The difference between the VFR(383K) values obtained under dry (12% RH) and humid conditions (50% RH) ranged from a few percent to 40%. In general, the greatest differences in VFR(383K) were observed at 288 K, and the water effect was less pronounced at 298 K. For most experiments, the use of 2-butanol as a scavenger produced greater VFR(383K) values than cyclohexane, which in turn yielded higher values than no scavenger at all.

For the highest and lowest relative humidities in each experiment, the volatility was measured at a number of evaporation temperatures from 298 K to 493 K. Representative plots of VFR against evaporation temperature under three sets of conditions used in this work (circles) are shown in Figure 3.

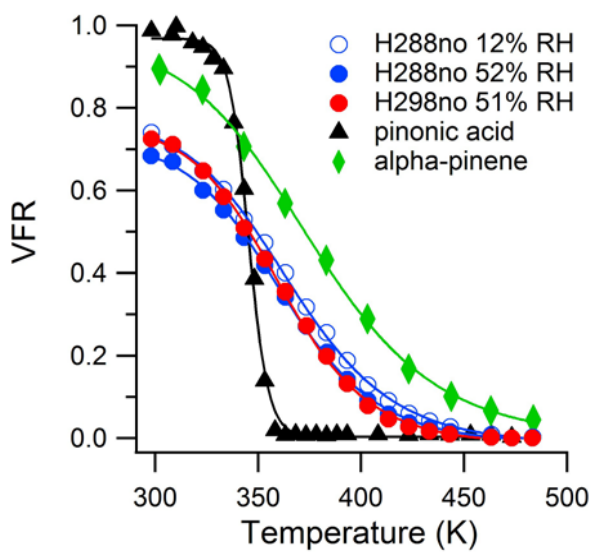

**Figure 3.** VFR versus evaporation temperatures for experiment H288no 12% RH (open blue circles), H288no 52% RH (closed blue circles) and H298no 51% RH (closed red circles) in this study. Solid lines show the sigmoidal fits to the data. Data for pinonic acid[10] (black triangles) and an α-pinene photo oxidation experiment in the SAPHIR outdoor chamber[15] (green diamonds) are included for comparison.

In general, volatility increased with the RH, as indicated by the lower VFR for H288no at 52% compared to 12% RH. The VFR values obtained in the two humid experiments, H288no and H298no, exhibited some difference depending on the evaporation temperature: when using low evaporation temperatures (< 383K), the high evaporation temperature (H298no) produced higher VFRs. However, there were no significant differences in VFR at higher evaporation temperatures.

One of the constraints encountered when using only one evaporation temperature is illustrated in Figure 3, in which it is clear that the shape of the curves depends on the experimental conditions. A sigmoidal function (equation 1) was fitted to the experimental data to describe the changes in VFR with evaporation temperatures.

Figure 3 shows both the measured data and the corresponding sigmoidal fits. The fits were used to obtain values for the parameters $T_{VFR0.5}$, $S_{VFR}$, $VFR_{298K}$ and $VFR_{523K}$ at high and low relative humidity for each experiment. The values of $T_{VFR0.5}$, $S_{VFR}$ and $VFR_{298K}$ for each experiment, except those in which the SOA number was too low for VTDMA analysis, are presented in Table 2. For the β-pinene experiments shown in Figure 3, $T_{VFR0.5}$ ranged from 342.2 to 349.3K and $S_{VFR}$ from -16.6 to -14.4. Data for pure nebulized pinonic acid[10] and a photo-oxidized aged α-pinene SOA[15] are also presented. The nebulized pinonic acid has a well-defined evaporation $T_{VFR0.5}$ of 345.3 K, and a steep slope ($S_{VFR}$) of -72.4, reflecting the narrow volatility distribution expected for a pure compound. The aged α-pinene has the highest $T_{VFR0.5}$ (372.9 K) and also the shallowest slope ($S_{VFR}$ = -12.2), implying a less volatile aerosol and a wider volatility distribution for the compounds within these particles. This was also the only experiment that yielded a residual at 523K ($VFR_{523K}$ = 0.012).

As can be seen in Figure 3, the three experimental conditions applied when using β-pinene produced different VFR values over the evaporation temperature range investigated. This is reflected in factors such as the extracted parameter $VFR_{298K}$, which increases in the order H288no (52% RH) < H298no (51%RH) ≤ H288no (12% RH). Thus, the semi-volatile fraction accounts for a greater proportion of the aerosol at high humidity and in the low temperature experiments. The slope of the plot ($S_{VFR}$) becomes steeper at high temperatures and high relative humidities.

Figure 4a shows the difference between high and low humidity conditions for $T_{VFR0.5}$ and $S_{VFR}$, respectively. Higher humidity leads to a reduced $T_{VFR0.5}$ and a steeper slope, i.e. a decrease in the slope factor $S_{VFR}$. The highest $T_{VFR0.5}$ value (356.9 K) was observed in the L288cyc experiment at low RH (11.6%), while the lowest was observed in the I288no experiment at high RH (49.4%). As shown in Figure 4, $T_{VFR0.5}$ is generally higher at low RH than at high RH. At 298 K, $T_{VFR0.5}$ is higher when using an OH-scavenger, and cyclohexane yields the highest $T_{VFR0.5}$ values while at 288 K this is valid for most experiments at low RH, Figure 4c. The volatility distribution is described by the slope factor $S_{VFR}$, which ranges from a gentle slope of -14.4 for H288no at low RH (12.8%) to the steepest slope of -18.9 for H288cyc at high RH (51.2%). $S_{VFR}$ is generally steeper at 298 K than at 288 K.

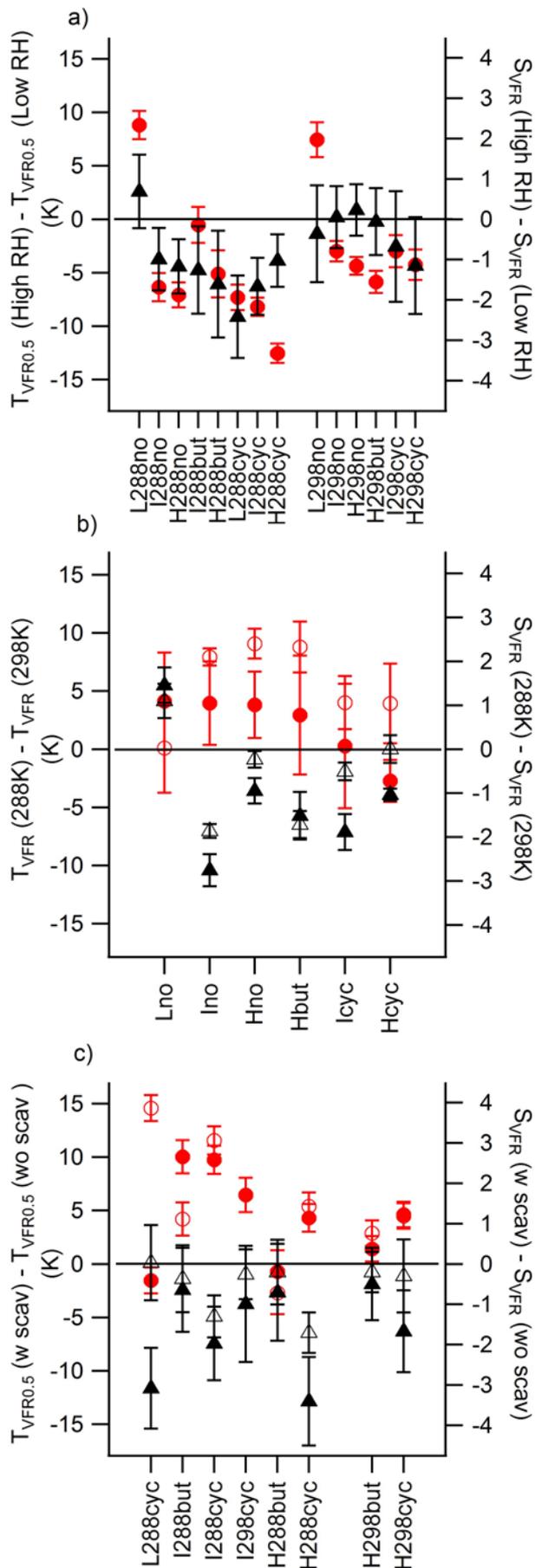

**Figure 4.** The differences in $T_{VFR0.5}$ (red circles, left vertical axis) and $S_{VFR}$ (black triangles, right vertical axis) between a) high and low relative humidity, b) high and low temperature and c) with and without scavenger. Filled symbols ~ 50% RH, open symbols ~ 12% RH.

At 288 K, the $S_{VFR}$ is always steeper (indicating a less complex composition of the aerosol) at high relative humidities (~ 50%) than at low RH (~ 12%). However, this relationship becomes less pronounced at 298 K (see Figure 4). The increase in humidity thus favors fewer reaction pathways while low humidities permit more competition, resulting in compounds with a larger distribution in vapor pressures. The slope is always steeper when using an OH scavenger, with cyclohexane giving the steepest slopes. Thus, higher OH radical availability makes the β-pinene chemistry more complex. In conjunction with the ongoing ozonolysis, this generates compounds with a larger distribution of vapor pressures. The difference between 2-butanol and cyclohexane should reflect the difference in the $HO_2$ to $RO_2$ ratios obtained with the two scavengers. In contrast to $VFR_{298K}$ and $T_{VFR0.5}$, the $S_{VFR}$ correlates with the number ($N_{10-300nm}$) and mass ($M_{10-300nm}$) concentrations of the produced SOA. When the slope is steeper, lower $M_{10-300nm}$ and $N_{10-300nm}$ values were obtained. It is not clear whether the slope becomes less steep due to the greater amount of SOA produced or whether a wider volatility distribution generates a larger $M_{10-300nm}$ and $N_{10-300nm}$.

The parameter $VFR_{523K}$ is an indicator of the amount of SOA remaining at high temperatures, and may be related to the progress of reactions that produce macromolecules or highly oxygenated products with very low vapor pressures. For all β-pinene experiments, the derived $VFR_{523K}$ was close to zero (< 0.05). We therefore conclude that the SOA produced in these experiments was fresh and did not include any low volatility macromolecules that are characteristic of aged SOA.[51]

**Discussion**

The negative water dependence and lower amounts of SOA produced as the $HO_2/RO_2$ ratio increased seem to contradict the effects observed in previous experiments on the ozonolysis of α-pinene, $\Delta^3$-carene and limonene using the G-FROST facility.[40,41] However, these three compounds all have an endocyclic double bond. Limonene has both an endo- and an exocyclic double bond, but it is likely that the initial ozone addition will predominantly occur at the endocyclic double bond.[43,52]

Ozonolysis of an exocyclic double bond such as that in β-pinene leads to the formation of two fragments. This will create smaller moieties in which the excited Criegee intermediate (CI*) will contain fewer oxygen atoms. The energy released in the decomposition of the POZ from an exocyclic compound will be partially distributed as kinetic energy over the two fragments.[30] For β-pinene, the dissociation of the POZ will generate nopinone or HCHO and one of two biradicals, $C_1$-CI* or $C_9$-CI* (see Scheme 1, R1 a and b,). The corresponding decomposition for α-pinene will lead to one aldehyde- or ketone-substituted biradical, $C_{10}$-CI*. The potential production of a secondary ozonide (SOZ) from the initially formed CI*, depends on the rate of stabilization and the available carbonyl species, which dictate the rate of reaction 3 b. For β-pinene, large amounts of nopinone and HCHO are produced (R1 and R3 a), while for α-pinene, the carbonyl group remains within the CI* moiety.

The effect of water on SOA formation has been attributed to a shift in chemical reaction paths[40], although observed mass changes could be caused by physical water uptake (if increased mass) or changes in the partitioning of organic compounds.[53] One obvious chemical explanation is seen in the mechanism shown in scheme 1, where water is directly involved in a reaction with the SCI that produces nopinone and $H_2O_2$ (R3 a).[23,31] The observed increase in the abundance of nopinone and $H_2O_2$ at higher relative humidities demonstrates that there is competition between the water reaction and other reaction(s) such as the formation of the SOZ (R3 b).[31] For SCIs with an α-hydrogen such as that produced in α-pinene ozonolysis, there is an alternative mechanism in which the water reaction produces an acid (see e.g. α-R3 c).[54,55]

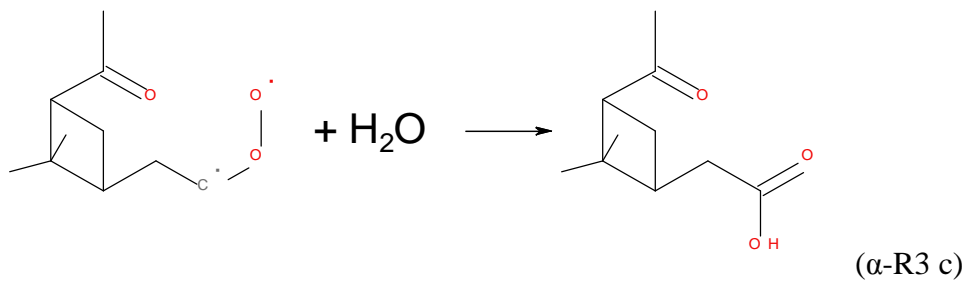

(α-R3 c)

Nopinone was one of the major products observed in the β-pinene ozonolysis experiments.[31] Being a volatile organic compound, it should not directly contribute to the mass or number of SOA particles. Therefore, if water promotes nopinone formation at the expense of other SOA-forming processes, the total SOA mass and number would decrease with increasing water concentrations. However, both nopinone and HCHO, which is the other major product of POZ dissociation via R1 b (HCHO), are carbonyls that can react with the available SCIs (R3 b) and form secondary ozonide, i.e. $C_{18}$-SOZ and $C_{10}$-SOZ.

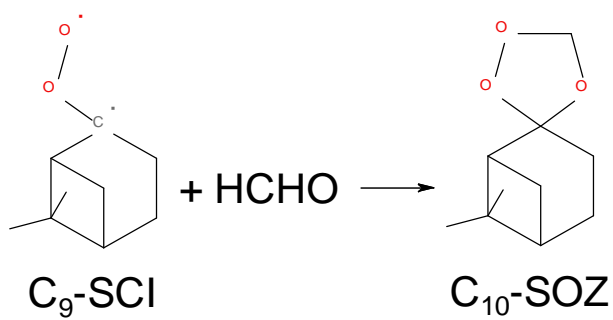

(R3 b1)

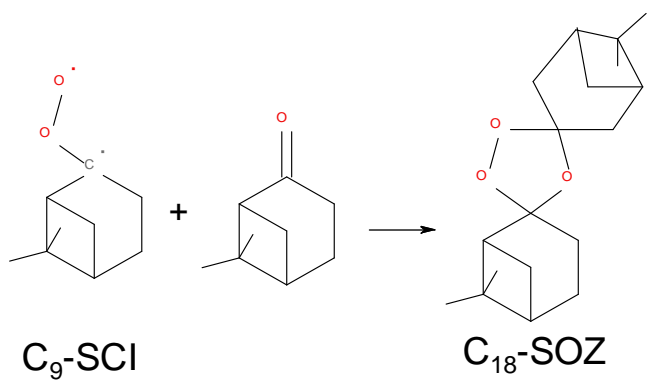

(R3 b2)

The $C_{18}$-SOZ has a low vapor pressure and could facilitate nucleation and contribute to SOA mass.[46] The production of $C_{18}$-SOZ thus explains the observed negative effect of water on SOA formation from β-pinene ozonolysis. Interestingly, a positive water effect on SOA mass was observed in α-pinene.[40] As discussed above, α-pinene can produce pinonic acid by reaction α-R3 c.[56] α-pinene ozonolysis is also less likely to produce secondary ozonides from bimolecular reactions since no closed shell carbonyl species are generated in the initial decomposition of the POZ. Moreover, the potential internal rearrangement of the SCI to produce the rather complex $C_{10}$-SOZ (α-R3 b) has yet to be demonstrated:

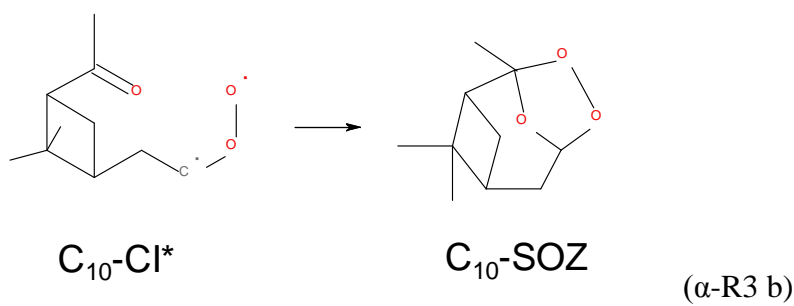

$C_{10}$-CI*   $C_{10}$-SOZ    (α-R3 b)

Thus, in the ozonolysis of α-pinene, the low volatility compound pinonic acid is produced from the alternative water channel (α-R3 c). However, if $C_{10}$-SOZ were produced in the competing reaction (α-R 3b), it would be more volatile, in keeping with a positive water effect on SOA formation.

Using data from a thermal desorption particle beam mass spectrometer, Docherty and Ziemann[23] concluded that the SOZ (R3 b) from β-pinene did not contribute to SOA formation. Instead they discussed the potential for water to influence the $HO_2$ to $RO_2$ ratio, which is known to affect SOA formation.[33] One potential fate of the alkyl radical produced in the hydroperoxide channel (R2 c) is radical propagation, which involves a number of $RO_2$ + $RO_2$ reactions that has a high yield of RO, e.g. 90% for the peroxy radical from β-pinene according to Master Chemical Mechanism (MCM) v3.[33] For each $RO_2$ + $RO_2$ reaction, there is a competing HO$_2$ reaction that may terminate the radical chain by forming a hydroperoxide or an acid as illustrated by reaction scheme 2.

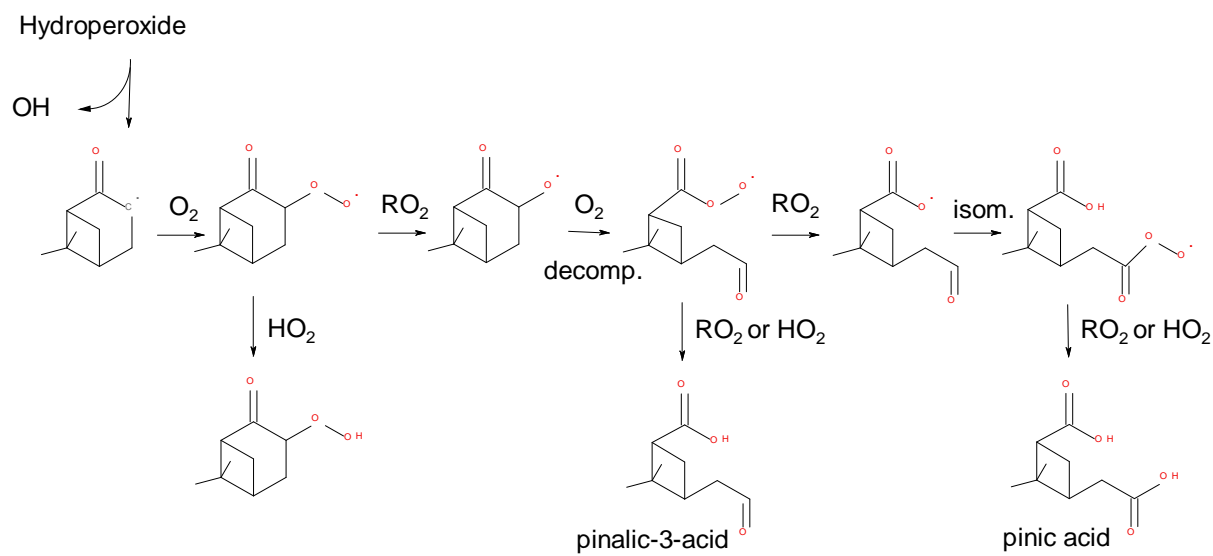

***Scheme 2.*** Formation of multifunctional oxygenated products by a sequence of radical reactions of β-pinene, starting with the alkyl radical from the hydroperoxide channel shown in scheme 1.

The effect of [HO$_2$]/[RO$_2$] was demonstrated for the ozone-initiated oxidation of β-pinene and α-pinene using the MCM.[33] In simulations using 2-butanol as the OH-scavenger (which would produce a high [HO$_2$]/[RO$_2$] ratio), enhanced concentrations of multifunctional products containing hydroperoxy groups were noted. It is not clear what happens to such hydroperoxy compounds, but it was suggested that they can react with carbonyls to generate peroxyhemiacetals.[6] The formation of peroxyhemiacetals from monoterpenes was subsequently observed by Docherty et al.[39] These authors concluded that for the exocyclic monoterpenes, an elevated [HO$_2$]/[RO$_2$] ratio increases the potential for forming highly volatile hydrogen peroxides at the expense of less volatile hydrogen peroxides.

A higher volatility makes the hydroperoxide less amenable to uptake by particles and reduces the potential for subsequent heterogeneous reaction with carbonyl compounds to form peroxyhemiacetals. This would explain the scavenger effect on SOA formation for β-pinene as observed in our study. However, the water effect in this case would most likely originate from the $HO_2$ self-reaction, which would reduce [$HO_2$] at high humidities. This would in turn decrease the $HO_2/RO_2$ ratio and thus counteract the consistently negative effect of water on SOA formation in β-pinene ozonolysis. Thus, the positive water effect for α-pinene can be explained by the $HO_2/RO_2$ ratio, but not the negative effect for β-pinene.

Keywood et al.[38] presented a mechanism of SOA formation involving acyl peroxy radicals, which could explain the difference between α-pinene and β-pinene. Using a simplified chemical mechanism, these authors showed that acid production, which can serve as an indicator of SOA formation, was sensitive to $HO_2/RO_2$ ratios. Simulations demonstrated higher acid yields when using cyclohexane instead of 2-butanol as the OH scavenger. This is consistent with our findings on SOA formation from β-pinene ozonolysis. In a second set of simulations, Keywood et al. incorporated a source of acyl peroxy radicals (5% of reacted precursor) via the hot acid/ester channel (R2 b). This changed the predicted production of acids, and 2-butanol as scavenger produced more acids than did cyclohexane, which is consistent with our previous work on α-pinene.[41] The mechanistic rationale is that the α-pinene CI* has an α-hydrogen that may produce the hot acid (α-R2 b), while the CI* from β-pinene would instead produce esters such as the lactones shown in scheme 1.

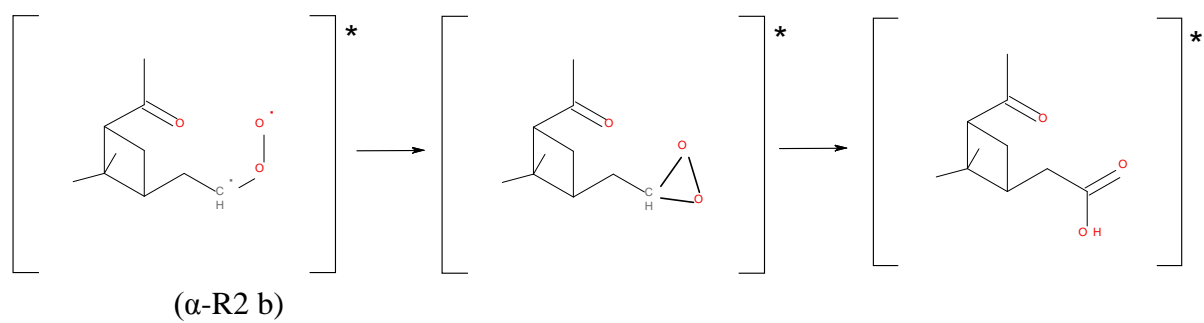

(α-R2 b)

For both α- and β-pinene, the traditional routes of acid production are from the acyl peroxy radicals generated in the hydroperoxide channel (scheme 2).[33,57] Ma et al.[57] identified and quantified a number of acids from α-pinene and β-pinene at different humidity and scavenger conditions. Based on the similarity of the observed responses, they concluded that the acids from α-pinene and β-pinene ozonolysis are probably produced via the same reaction channel, e.g. the hydroperoxide channel. This is somewhat inconsistent with the additional formation of acyl peroxy radicals from the α-pinene system via the ester channel as suggested by Keywood et al.[38]

The combined water and scavenger effects observed in the present work cannot be fully explained by the mechanisms discussed above. There appears to be missing chemistry that would account for the water dependence for both endocyclic and exocyclic compounds. In an investigation into OH radical formation from alkene ozonolysis and its water dependence, Anglada et al.[55] suggested that the hydroxyl hydroperoxide generated by the water reaction of SCI (R3 d) can be decomposed, and that there is a water-catalyzed channel (R3 e):

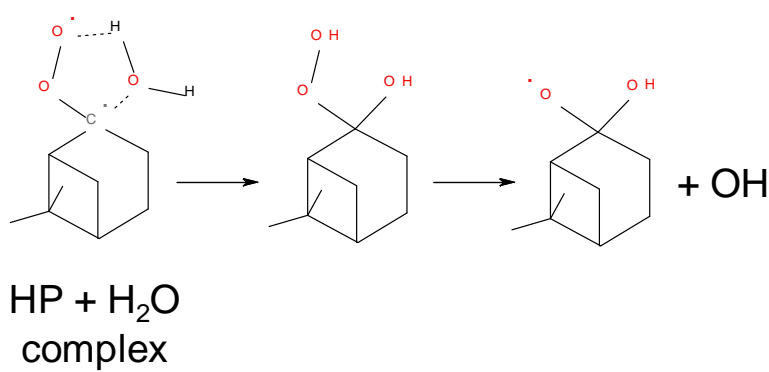

HP + H₂O complex

(R3 d)

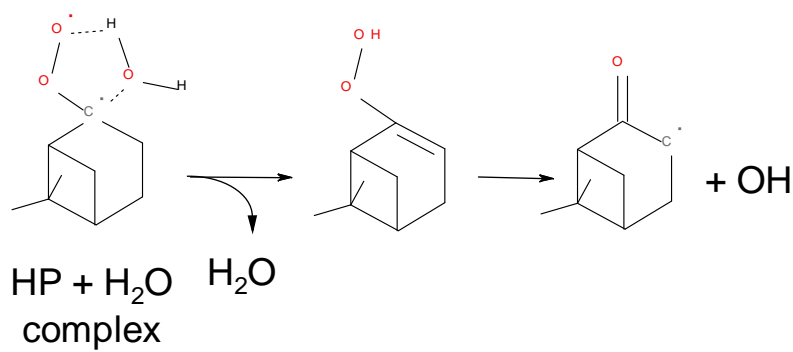

(R3 e)

In the case of β-pinene, ring strain might limit the viability of the water-catalyzed reaction. However, in α-pinene, this process could create a carbonyl group while maintaining the radical propagation that creates multi-functional oxygenated products. As outlined, this process and reaction α-R3 c would account for distinct differences in the products of the water reactions between the SCIs produced by endocyclic compounds, such as α-pinene, and the exocyclic β-pinene.

Finally, the suggestion by Jonsson et al.[42] involving the hydroxyl hydroperoxide from α-pinene, acyl peroxy radical chemistry[38] and the OH formation pathway R3 d[55] (scheme 3).

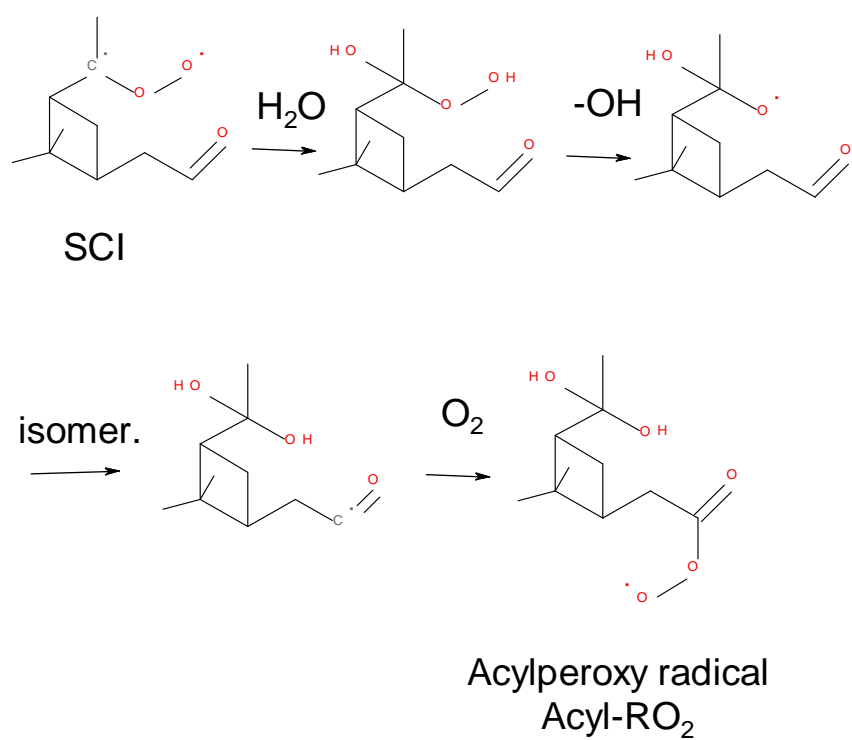

*Scheme 3.* Formation of acylperoxy radicals via a water reaction with the α-pinene SCI.

In this suggestion, the final step is an internal H-abstraction from the aldehyde group. Whether this specific H-migration would be favored compared to other positions within the alkoxy radical, is an open question since the lower bond strength for the aldehyde hydrogen is counteracted by other factors.[58] However, this path is impractical for the corresponding alkoxy radical produced from β-pinene. Instead, other H-migrations may be possible even if they should be limited by the radical's bicyclic structure. This again illustrates that β-pinene has less scope for forming low volatility products from the water reaction with the stabilized CI, which may explain the difference between the two compounds in terms of the relationship between water and SOA formation. Furthermore, if the decomposition routes (R2 b and/or R2 c) compete with any other water dependent channel (such as R3 a, c, d and e), the link between the chemical mechanism and our experimental findings would be straightforward. Indeed, recent studies have suggested that the stabilized CI can also rearrange into a hydroperoxide (R3 f).[30,59,60]

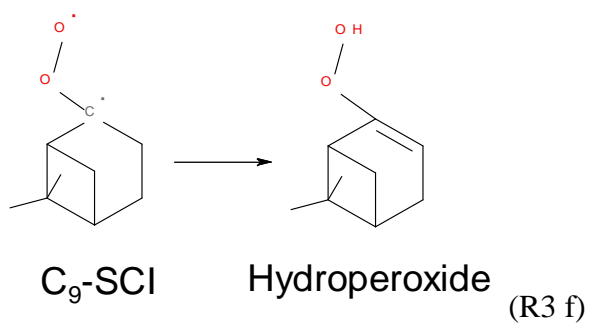

$C_9$-SCI      Hydroperoxide
                              (R3 f)

The mechanism that forms this hydroperoxide is analogous to R2 c, but will directly compete with the water reactions R3 a, c-f. For β-pinene, nopinone formation (R3 a) will then be in competition with the generation of multi-functional oxygenated products via the subsequent $HO_2/RO_2$ chemistry of the hydroperoxide channel (R3 f). The lifetime of the SCI from β-pinene with respect to H-migration (R3 f) is 0.3 s.[59] This is comparable to that for the water reaction if one assumes a bimolecular rate coefficient of 4.3 x $10^{-19}$ $cm^3$ $molecule^{-1}$ $s^{-1}$ as proposed by Nguyen et al.[30] . Under these conditions, the lifetime of the SCI would be ~5 s at 50% RH. In addition, one must consider the possibility of a water dimer reaction, which has been suggested to be even faster than the water reaction under atmospheric conditions.[61] A similar effect on water dependence would be seen if water could react directly with the CI* (R2 e) or were more efficient in quenching CI* to form the SCI (R2 a).

The results from the experiments at 288 K follow the trends observed at 298 K. This is somewhat surprising given that the absolute water concentration differs by a factor of two between these two temperatures. In principle, this should shift the point at which the water reaction becomes competitive with other channels to a higher relative humidity, and the humidity effect should be smaller at low temperatures. Our data indicate that raising the temperature reduces the low temperature humidity effect, but only with 20% on average (ranging from 2 up to 60%) compared to the doubling in absolute water content.

The volatility data could partly be understood from the mechanism described above. The effects on volatility from scavengers and humidity were similar at the two temperatures studied. This can partly be caused by the SOA produced at 288 K being warmed to 298 K before entering the first DMA in the VTDMA equipment. This will attenuate the effect of partitioning in favor of the effect of chemistry on the SOA properties. In keeping with the mechanism the slope factor $S_{VFR}$ at both temperatures decreased in the order no scavenger > 2-butanol > cyclohexane (Figure 4). This could be linked to the more complex chemistry that occurs in the absence of a scavenger, and may also indicate that the additional $HO_2$ chemistry in the 2-butanol case gives products with a wider range of vapor pressures. Increases in relative humidity gave lower $T_{VFR0.5}$ values and a steeper slope. This shows that routes that producing less volatile material become negligible when the water reaction is increased. This was reflected in an overall volatility increase, and by a reduction in the range of vapor pressures for the compounds contributing to the aerosol. This is in line with the lower mass and number concentrations of the SOA produced at high relative humidities.

**Conclusions**

The new data on the influence of RH, temperature, initial rate and scavenger conditions on SOA mass and number formation from ozonolysis of β-pinene, clearly show how it differs from the analogous α-pinene system. Several literature-based explanations for these differences have been suggested. These explanations have been used to describe the simultaneous effects of water and radical conditions, where increases in RH also enhance the scavenger effect on the mass and number of SOA particles. Furthermore, the magnitude of the humidity effect on particle numbers decreased as the concentration of the precursor β-pinene increased, and the effect of humidity was more pronounced when using an OH scavenger. The most reasonable explanation for this combined effect is the recently suggested decomposition of the stabilized CI into hydroperoxide.[30,59,60] The measured evaporation at a range of temperatures was parameterized by determining $VFR_{298K}$, $T_{VFR0.5}$, $S_{VFR}$ and $VFR_{523K}$. The introduction of experimental proxy volatility parameters that reflect important aspects of volatility of SOA is well warranted. It was demonstrated that these parameters could be used in a detailed evaluation of the experimental conditions, and be linked to changes in the chemical mechanism of ozonolysis.

**Acknowledgement**

The research presented is a contribution to the Swedish strategic research area ModElling the Regional and Global Earth system, MERGE. This work was supported by Formas (214-2010-1756), the Swedish Research Council (80475101) and the Swedish EPA research programme CLEO. EUE acknowledges support from the platform initiatives at the Faculty of Science,

University of Gothenburg. Dr. Johan Engelbrektsson is acknowledged for the LabVIEW control of G-FROST.

**Table of Contents Image**

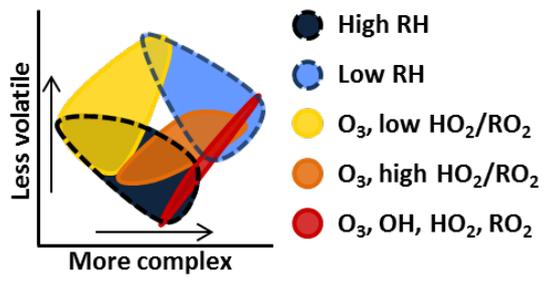